\documentclass[twocolumn,
		]{revtex4}
\usepackage{graphicx,color}

\usepackage{amssymb}
\usepackage{amsbsy}
\usepackage{amsthm,mathrsfs,amsopn}
\usepackage{amsmath}

\theoremstyle{plain} 

\begin{document}

\title{Preferential attachment with partial information}

\author{Timoteo Carletti$^{1}$, Floriana Gargiulo$^{1}$, Renaud Lambiotte$^{1}$}
\affiliation{
1. Department of Mathematics and Namur Center for Complex Systems - naXys, University of Namur, rempart de la Vierge 8, B 5000 Namur, Belgium}

\begin{abstract} 
We propose a preferential attachment model for network growth where new entering nodes have a partial information about the state of the network. Our main result is that the presence of bounded information modifies the degree distribution by introducing an exponential tail, while it preserves a power law behaviour over a finite small range of degrees. On the other hand, unbounded information is sufficient to let the network grow as in the standard Barab\'asi-Albert model. Surprisingly, the latter feature holds true also when the fraction of known nodes goes asymptotically to zero. Analytical results are compared to direct simulations.
\end{abstract}

\maketitle

\section{Introduction}
\label{sec:intro}

Since the pioneering works of Erdos-Reny~\cite{ER1959,ER1960} and Rapoport~\cite{Rapoport1957}, network theory has become a central topic of research, providing a set of tools and algorithms adapted to any type of system made of elements in interaction, with applications in a broad range of scientific disciplines. The increasing availability of large-scale data in networked systems has lead, over the last 15 years, to a new wave of research \cite{Rev}, with the identification of universal properties in social, biological and information systems \cite{WS,BA1999,Faloutsos}, the development of algorithms to extract information from their structure \cite{Fortunato}, and the analysis of theoretical models reproducing the observed patterns and explaining the impact of structure on dynamics \cite{Newman}.

A key property of complex networks is their broad degree distribution, often described in terms of power-laws, -- an observation not always supported by statistical evidence \cite{Stumpf}--, with important implications in terms of resilience and dynamics. As is often observed, the degree distribution strongly deviates from a binomial distribution, and presents a high heterogeneity, as most of the nodes have very few connections and few of them act as well-connected hubs. A well-known model for generating scale-free networks is the {\em preferential attachment} (PA) model of Barab\'asi-Albert ~\cite{BA1999}, where the network is assumed to grow, and where new nodes preferentially connect to highly connected nodes, with a probability proportional to their degree. This model, and variants of it, are known to asymptotically produce networks where the tail of the degree distribution behaves like $p_k\sim k^{-\alpha}$, where $\alpha$ depends on the details of the model. 

PA suffers from a series of limitations, such as its lack of local motifs and community structure, but also its unrealistic assumptions. First, PA only focuses on the role of degree and neglects other types of constraints for link formation. Important examples include the effects of ageing, as nodes tend to lose their ability to acquire new links as time goes on \cite{Age,Lambiotte2007}, of physical distance in spatial networks \cite{Barthelemy,Space}, or of homophily in social networks \cite{McPherson}. These additional constraints tend to limit the effective size of the system when a new node enters it, and to introduce cut-offs to the power-laws generated by the models, as observed in empirical data \cite{Clauset}. Another important limitation is the global nature of PA, as a new node
requires complete knowledge of the degrees of all existing nodes to decide which connections to draw. However, it is a piece of information not usually available at nodes in real systems \cite{Vazquez}. In order to circumvent this limitation, local models of network growth have been proposed, such as redirection, copying, duplication, or local exploration by random walks\cite{Vazquez,Valverde,Ispolatov,Evans}.

The main purpose of this paper is to explore further  the effects of partial, local information on network growth. Our model assumes that a node, when entering the system, only has access to a fraction of the existing nodes, and that it creates new connections with nodes inside this known set (KS) with a probability proportional to their degree. By construction, PA is thus applied, but only to a subset of the whole system. Let us observe that our model is reminiscent of, but different from, the one proposed in~\cite{SKM2007} where  the new node connects to the node with highest degree in KS. 
We show that, if the size of KS steadily increases with time, and hence with network size, the network asymptotically exhibits the same degree distribution as in the original PA model. Surprisingly, this result holds true even if the fraction of nodes in KS goes to zero as the network size increases. On the other hand, if the size of KS is bounded, and in particular if it is a constant, the asymptotic distribution converges to power law with an exponential cut-off, that can be characterised in terms of this maximum value. The analytical results are complemented and supported with dedicated numerical simulations.

\section{The model}
\label{sec:model}

The network $G_0$ is initially composed by a single node. At each time step, a new node enters the system and creates a new link with an already existing node~\footnote{One could generalize the process such that $H$ new nodes enter the system and make $K\geq H$ new links.}. In general, the latter is selected proportionally to its degree among $M_t\geq 1$ nodes randomly selected in the network, where $t$ is the time step. This set of $M_t$ randomly selected nodes is KS defined in the introduction and it is renewed at each time step. The parameter $M_t$ is a measure of the information held by the entering node, and it can, in general, vary during the network growth. It is an ingredient of the model, and we will focus on different scenarios for its evolution in the following.
We denote by $s=0$ the single node in $G_0$. Each subsequent node is labelled by the timing $s=t$ of its introduction in the network. The number of nodes at time $t$ is given by $N_t=t+1$ and the number of links $E_t=t$. Let us denote by $z_t^s$ the degree of node $s$ at time $t$, such that we have $s\leq t$ and trivially $\sum_{s=0}^t z^s_t=2t$.

The time evolution of the probability 
 $q_{t}(s,k)$ 
that node $s$ has degree $k$ at time $t$ is given by
\begin{eqnarray}
\label{eq:memicro}
q_{t+1}(s,k)=\frac{M_t}{N_t}\frac{k-1}{\Sigma_t}q_t(s,k-1)+\left(1-\frac{M_t}{N_t}\right)q_t(s,k) \cr
+\frac{M_t}{N_t}\left(1-\frac{k}{\Sigma_t}\right)q_t(s,k)\, .
\end{eqnarray}
The normalizing factor is given by $\Sigma_t=\sum_{j\in KS_t}^t z^j_t$, where $KS_t$ is the known set at time $t$ composed of the $M_t$ randomly selected nodes. The terms on the right hand side denote respectively: the probability that $s$ has degree $k-1$, it has been selected in KS and the new link is established with it proportionally to its degree. The second term represents the probability that $s$ has already degree $k$ and it does not belong to KS; finally, the last term denotes the probability that $s$ has degree $k$, it belongs to KS, but it has not been chosen to be linked with. Eq.(\ref{eq:memicro}) is complemented with the conditions
\begin{equation}
\label{eq:inicond}
q_1(0,k)=q_1(1,k)=\delta_{k,1} \quad \text{and}\quad q_t(t,k)=\delta_{k,1} \quad\forall t\geq 2\, ,
\end{equation}
namely the initial two nodes $s=0$ and $s=1$ have degree $k=1$ at time $t=1$ and the entering node always has degree $k=1$.

In the following, we are interested in the fraction of nodes that have degree $k$ at time $t$
\begin{equation}
\label{eq:probavn}
p_t(k)=\frac{1}{N_t}\sum_{s=0}^{t}q_t(s,k)\, .
\end{equation}
The time evolution of $p_t(k)$ is straightforwardly obtained by plugging its definition into the rate equation Eq.~\eqref{eq:memicro}
\begin{eqnarray}
\label{eq:me1stp}
N_{t+1}p_{t+1}(k)=\delta_{k,1}+M_t\frac{k-1}{\Sigma_t}p_t(k-1)+N_tp_t(k) \cr
-M_t\frac{k}{\Sigma_t}p_t(k) \, .
\end{eqnarray}

To go one step further, we need to estimate the normalizing factor $\Sigma_t$, that is the sum of the degrees of $M_t$ randomly selected nodes. First, let us observe that, among such nodes, we know that one node
has degree $k-1$ in the second term, while it has degree $k$ in the fourth term. Neglecting such nodes, the remaining sum in the normalizing factors is approximated by  
\begin{equation}
\label{eq:estnormfact}
\sum_{j\in KS^{\prime}_t}^t z^j_t=2t\frac{M_t-1}{N_t}\, ,
\end{equation}
that is $(M_t-1)$ times the average degree  $\bar{k}_t= \frac{2t}{N_t}$ in the network. This approximation is expected to be accurate when $N_t$ is sufficiently large, due to the central limit theorem, but it is also expected to be rough when $N_t$ is small, especially due to the expected heterogeneity of the degree distribution. However, the very good agreement between analytical results and numerical simulations support its validity in both cases. Under this approximation, the rate equation \eqref{eq:me1stp} rewrites as follows:
\begin{eqnarray}
\label{eq:me}
N_{t+1}p_{t+1}(k)
=\delta_{k,1}+N_tp_t(k)\cr
+M_t\frac{k-1}{(k-1)N_t+2t(M_t-1)}N_tp_t(k-1)\cr
-\frac{M_t k}{kN_t+2t(M_t-1)}N_tp_t(k)
\, .
\end{eqnarray}

\section{Asymptotic solution}
\label{sec:analsol}

Let us first observe that Eq.~\eqref{eq:me} contains two well-known results as limiting cases. If the information is at its smallest value, that is $M_t=1$, the model reduces to a random attachment scheme. At each time step, one node is chosen at random among the existing ones and a link is formed with the new node. In that case, Eq.~\eqref{eq:me} reduces  to
\begin{equation}
\label{eq:meexp}
N_{t+1}p_{t+1}(k)=\delta_{k,1}+N_tp_t(k)+p_t(k-1)-p_t(k)\, ,
\end{equation}
whose asymptotic solution $p^{(rnd)}_{\infty}(k):=\lim_{t\rightarrow
  \infty}p_{t}(k)$  is
\begin{equation}
\label{eq:meexp3}
p^{(rnd)}_{\infty}(k)=\frac{1}{2^{k}}\quad \forall k\geq 1\, ,
\end{equation}
and an exponential distribution is recovered. 

Another limit case is when $M_t$ takes its maximum value, $M_t=N_t$, and the model reduces to the standard PA model~\cite{BA1999} 
\begin{equation}
\label{eq:mepl}
N_{t+1}p_{t+1}(k)=\delta_{k,1}+N_tp_t(k)+(k-1)\frac{N_t}{2t}p_t(k)-k\frac{N_t}{2t}p_t(k)\, ,
\end{equation}
whose asymptotic solution is well-known to be
\begin{equation}
\label{eq:mepl3}
p^{(PA)}_{\infty}(k)=\frac{6}{k(k+1)(k+2)}\, ,
\end{equation}
and whose tail behaves like $k^{-3}$.

In order to compute the analytical solution of
Eq.~\eqref{eq:me} for generic choices of $M_t$, let us consider two distinct cases, when it is bounded and it is unbounded. We further make the realistic assumption that $M_t$ is a growing function of time~\footnote{Let us observe that it seems natural not to loose information as time goes on and thus $M_t$ to be monotone increasing, however mathematically one only needs the convergence of $M_t$ for $t\rightarrow \infty$ and thus more exotic functions $M_t$ will fit in our scheme.}, i.e. $M_{t+1}\geq M_{t}, \forall t$.

\subsection{Bounded information}
\label{eq:bounded}

Because it is growing, the bounded sequence $M_t$ asymptotically converges to the limit $M_{\infty}$ when $t$ is sufficiently large.
The asymptotic distribution
$p^{(bnd)}_{\infty}(k)$ is thus solution of
\begin{eqnarray}
\label{eq:mebounded}
p^{(bnd)}_{\infty}(k)=\delta_{k,1}+\frac{M_{\infty}(k-1)}{k-1+2(M_{\infty}-1)}p^{(bnd)}_{\infty}(k-1)\cr
-\frac{M_{\infty} k}{k+2(M_{\infty}-1)}p^{(bnd)}_{\infty}(k)\, ,
\end{eqnarray}
where we used the fact that $\bar{k}_t\rightarrow 2$.
Setting $\pi_k=kM_{\infty}p^{(bnd)}_{\infty}(k)$, we can rewrite the previous equation
as
\begin{eqnarray}
\label{eq:mebounded2}
\pi_k = \frac{kM_{\infty}
  (k+2(M_{\infty}-1))}{k(M_{\infty}+1)+2(M_{\infty}-1)}\frac{\pi_{k-1}}{k-1+2(M_{\infty}-1)},\cr
\forall k\geq 2\, ,
\end{eqnarray}
and thus
\begin{eqnarray}
\label{eq:mebounded3}
\pi_k = \frac{M_{\infty}(k+2(M_{\infty}-1))}{3M_{\infty}-1}\left(\frac{M_{\infty}}{1+M_{\infty}}\right)^{k-1}\cr
\frac{k!}{\prod_{j=2}^{k}\left[j+ 2(M_{\infty}-1)/(M_{\infty}+1)\right]}\quad
\forall k\geq 2\, ,
\end{eqnarray}
which leads to the following solution for  $p^{(bnd)}_{\infty}(k)$, in terms of the Euler Beta function $B(x,y)$~\footnote{Let us recall that the Euler Beta function is defined by $B(x,y)=\Gamma(x)\Gamma(y)/\Gamma(x+y)$, being $\Gamma(x)$ the Gamma function $\Gamma(x)=\int_0^{\infty}s^{x-1}e^{-s}\, ds$.} 
\begin{eqnarray}
\label{eq:mebounded4}
p^{(bnd)}_{\infty}(k) = \left(\frac{M_{\infty}}{1+M_{\infty}}\right)^{k-1}\frac{k+2(M_{\infty}-1)}{k(M_{\infty}+1)}\cr
B\left(k,\frac{2(M_{\infty}-1)}{M_{\infty}+1}\right)\quad
\forall k\geq 1\, .
\end{eqnarray}

%

\begin{figure}[h]
\begin{center}
\hspace*{-1cm}
\includegraphics[width=7cm]{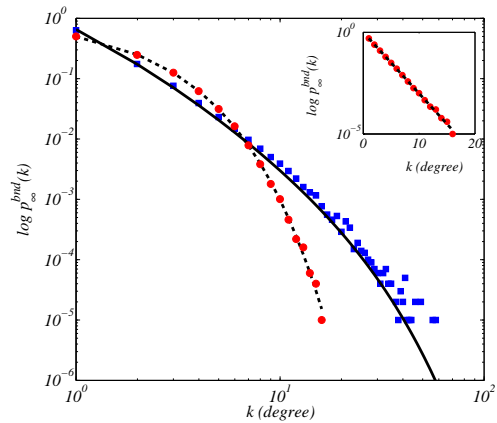}
\end{center}
\caption{Asymptotic probability distribution $p^{(bnd)}_{\infty}(k)$ for bounded $M_{\infty}$ (loglog plot). Main panel, solid lines correspond to the analytical solution, while dots to a numerical realisation of the network with $N_t=10000$ nodes, red (on line) circles with $M_{\infty}=1$ and blue (on line) squares with $M_{\infty}=10$. Inset, we report the case $M_{\infty}=1$ in semilogarithmic scale to appreciate the exponential law.}
\label{fig:llMbdb}
\end{figure}

As previously claimed, when $M_{\infty}=1$ the asymptotic distribution follows an exponential law, as shown in Fig.~\ref{fig:llMbdb}. For larger but still bounded $M_{\infty}$, the distribution presents two regimes: for small $k$ (compared to $M_{\infty}$), the distribution is close to a power law, while for large $k$ (still compared to $M_{\infty}$), it follows an exponential law (see Fig.~\ref{fig:llMbdb}). 

This behaviour can be understood by the following arguments. Let us take $M=M_{\infty}$ strictly constant for the sake of simplicity. If $M$ is larger than the size of the system, which is valid for small enough times, the entering node has a complete information about the system, and the model behaves like the standard PA. Moreover, because $t$ is small, the degrees of the nodes involved in this process are also small. When the size of the network is sufficiently large, in contrast, $M$ becomes too small to provide a good sampling of the network, and the new nodes attach almost randomly to the existing ones, which leads to an  exponential tail for the distribution. This exponential law dominates for large degrees, because it emerges at a time when the network is large enough to exhibit large degrees.

\begin{figure}[h]
\begin{center}
\hspace*{-1cm}
\includegraphics[width=7cm]{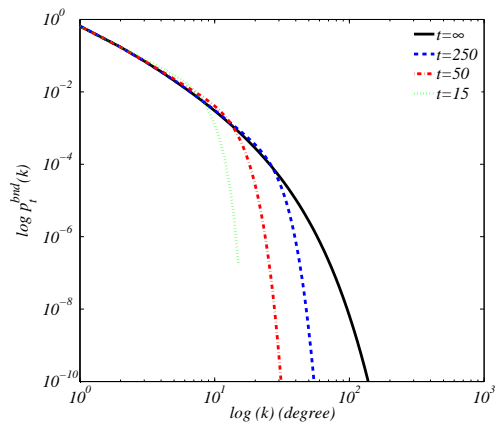}\quad
\end{center}
\caption{Time evolution of the degree probability distribution $p_t(k)$ for $M_{\infty}$ (loglog plot), for $M_{\infty}=10$. The solid black line corresponds to the asymptotic distribution, the dashed, blue (on line) curve to a large but finite time ($t=250$), the dot-dashed, red (on line) curve to an intermediate time ($t=50$) and the dotted, green (on line) to a very short time ($t=15$). One observes that the \lq\lq power law\rq\rq behaviour emerges for small times, and small degrees, and that the exponential tail is progressively filled as time goes on, until the asymptotic distribution is reached.}
\label{fig:pkt}
\end{figure}

To prove our claim more rigorously,  let us recall a basic property of the Euler Beta function:
\begin{equation*}
B(x,y)\rightarrow \Gamma(y)x^{-y}\quad \text{for }x\rightarrow \infty\quad\text{and $x>>y$.}
\end{equation*}
As $M_{\infty}$ is bounded, taking the limit of $k >> M_{\infty}$ leads to
\begin{eqnarray}
\label{eq:mebounded5}
\log p^{(bnd)}_{\infty}(k) \sim k\log \frac{M_{\infty}}{1+M_{\infty}}-2\frac{M_{\infty}-1}{M_{\infty}+1}\log k \cr
+\log \Gamma\left(\frac{2(M_{\infty}-1)}{M_{\infty}+1}\right)\, ,
\end{eqnarray}
and thus to an exponential behaviour for large $k$: 
\begin{equation}
\label{eq:mebounded5a}
p^{(bnd)}_{\infty}(k)\sim \alpha^{-k}\, ,
\end{equation}
with $\alpha =\frac{1+M_{\infty}}{M_{\infty}}\in(1,2]$.
Let us also recall that $B(x,y)\rightarrow 1$ for $y\rightarrow 0$, hence if $M_{\infty}\rightarrow 1$ we recover the exponential distribution of the random model, as expected:
\begin{equation}
\label{eq:mebounded7}
\lim_{M_{\infty}\rightarrow 1}p^{(bnd)}_{\infty}(k) = \frac{1}{2^{k}}\quad
\forall k\geq 1\, .
\end{equation}

In the other extreme, fixing $k$ and taking the limit $M_{\infty} >> k$, we get
\begin{equation}
\label{eq:mebounded6}
p^{(bnd)}_{\infty}(k) \sim \frac{2}{k}B\left(k,2)\right)=\frac{2}{k}\frac{\Gamma(k)\Gamma(2)}{\Gamma(k+2)}=\frac{2}{k^2(k+1)}\sim \frac{1}{k^3}\, .
\end{equation}

\subsection{Unbounded information}
\label{eq:unbounded}

Let us now consider the case when $M_t$ diverges with $t$, keeping in mind that it satisfies the bound $M_t\leq N_t$, as a node can not have more information than the total number of nodes at time $t$. By using the fact that $\bar{k}_t\rightarrow 2$, the following expression simplifies as
\begin{equation*}
\frac{kM_t}{k+(M_t-1)\bar{k}_t}\rightarrow \frac{k}{2}\, ,
\end{equation*}
which allows to rewrite Eq.~\eqref{eq:me} as follows:
\begin{equation}
\label{eq:meunbnb1}
p^{(unbnd)}_{\infty}(k)=\delta_{k,1}+\frac{k-1}{2}p^{(unbnd)}_{\infty}(k)-\frac{k}{2}p^{(unbnd)}_{\infty}(k)\, .
\end{equation}
This equation is exactly equivalent to Eq.~\eqref{eq:mepl}, originally obtained for the standard PA, which allows us to deduce that the dynamics is not affected by the partial amount of information in the case when this information is growing and unbounded. 

\begin{figure}[h]
\begin{center}
\hspace*{-1cm}
\includegraphics[width=6cm]{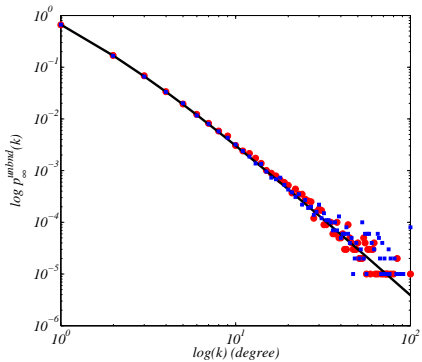}
\end{center}
\caption{Asymptotic probability distribution $p^{(unbnd)}_{\infty}(k)$ for unbounded $M$ (log log plot). The solid line correspond to the analytical solution~\eqref{eq:mepl3}, while dots correspond to a numerical realisation of the network with $N_t=10000$ nodes, blue (on line) squares with $M_t=t/10$ and red (on line) circles with $M_t=\sqrt{t}$.}
\label{fig:Munbdb}
\end{figure}

Fig.~\ref{fig:Munbdb} shows the very good agreement (and thus {\em a posteriori} the goodness of the approximation~\eqref{eq:estnormfact}) between the analytical results and numerical simulations of the model. Two different growth behaviours for $M_t$ are reported in the Figure, the case $M_t=t/10$ (blue squares), corresponding to a finite fraction, $M_t/N_t=1/10$, of nodes that are accessible, and the case $M_t=\sqrt{t}$ (red circles), where the fraction of known nodes goes to zero as the network size increases, i.e. $M_t/N_t\rightarrow 0$.

\section{Conclusion}
\label{eq:conc}

In this paper, we have considered a variation of the well-known preferential attachment model for network growth. Our main purpose was to explore the effects of partial information on network dynamics. We have shown that the presence of bounded information modifies the degree distribution by introducing an exponential tail, while it preserves a power law behaviour for small $k$, thus over a finite range of degrees. In the case of unbounded information, in contrast, the network grows as in the standard PA model. Surprisingly, this property also holds true also in situations when the fraction of known nodes goes to zero. 

This model offers an interesting explanation for the emergence of power-laws with cut-offs, as generally observed in empirical data, without requiring a finite capacity for the nodes, but by emphasising instead the imperfect sampling of the network when new nodes enter the system.

\section*{Acknowledgments}
This work presents research results of the Belgian Network DYSCO (Dynamical Systems, Control, and Optimization), funded by the Interuniversity Attraction Poles Programme, initiated by the Belgian State, Science Policy Office. The scientific responsibility rests with its author(s).


\begin{thebibliography}{99}


\bibitem{ER1959} O. Erd\H{o}s and A. R\'enyi, Publicationes Mathematicae \textbf{6}, 290‚Äì297 (1959)

\bibitem{ER1960} O. Erd\H{o}s and A. R\'enyi,  Publications of the Mathematical Institute of the Hungarian Academy of Sciences \textbf{5}, 17‚Äì61 (1960)

\bibitem{Rapoport1957} A. Rapoport, Bulletin of Mathematical Biology \textbf{19}, 257-77 (1957)

\bibitem{Rev} M. Newman, A.-L. Barab\'asi, and D.J. Watts, The Structure and Dynamics of Networks, 

\bibitem{WS} D.J. Watts and S. H. Strogatz, Nature \textbf{393}, 440-442 (1998)

\bibitem{BA1999} A.-L. Barab\'asi and R. Albert, Science \textbf{286}, pp. 509 (1999)

\bibitem{Faloutsos}
M. Faloutsos, P. Faloutsos and C. Faloutsos,  Comput. Commun. Rev. \textbf{29}, 251–263 (1999)

\bibitem{Fortunato}
S. Fortunato. Phys. 
Rep. \textbf{486}, 174 (2010) 

\bibitem{Newman}
M.E.J. Newman, SIAM Review \textbf{45},167--256 (2003)

\bibitem{Stumpf}
M.P.H. Stumpf, and M.A. Porter, Science \textbf{335}, 665-666 (2012)

\bibitem{Age}
K.B. Hajra and P. Sen, Phys. Rev. E \textbf{70}, 056103 (2004)

\bibitem{Lambiotte2007} 
R. Lambiotte, J. Stat. Mech., P02020 (2007)

\bibitem{Barthelemy}
Marc Barthelemy, Physics Reports \textbf{499},1-101 (2011)

\bibitem{Space}
R Lambiotte, VD Blondel, C De Kerchove, E Huens, C Prieur, Z Smoreda and P Van Dooren, Physica A \textbf{387}, 5317--5325 (2008)

\bibitem{McPherson}
J.M. McPherson, L. Smith-Lovin and J.M. Cook, Annu Rev Sociol \textbf{27}, 415--444 (2001)

\bibitem{Clauset}
A. Clauset, C.R. Shalizi, and M.E.J. Newman, SIAM Review \textbf{51}, 661-703 (2009).

\bibitem{Vazquez}
A. V\'azquez, Phys. Rev. E \textbf{67}, 056104 (2003) 

\bibitem{Valverde}
S. Valverde and R.V. Sol\'e, Europhys. Lett. \textbf{72}, 858 (2005)

\bibitem{Ispolatov}
I. Ispolatov, P.L. Krapivsky and A. Yuryev, Phys. Rev. E \textbf{71}, 061911 (2005)

\bibitem{Evans}
T.S. Evans and J. Saram\"aki,  Phys.Rev.E \textbf{72}, 26138 (2005) 

\bibitem{SKM2007} 
R.M. D'Souza, P.L. Krapivsky and C. Moore, The European Physical Journal B \textbf{59}, 535--543 (2007)



\end{thebibliography}
\end{document}